# Fabrication of Periodic, Flexible and Porous Silicon Microwire Arrays with Controlled Diameter and Spacing: Effects on Optical properties

Anjali Saini [1,2], Mohammed Abdelhameed [3,4], Divya Rani [5], Wipakorn Jevasuwan [3], Naoki Fukata [3], Premshila Kumari [1,2], Sanjay K. Srivastava [1,2], Prathap Pathi [1,2], Arup Samanta [5,6] and Mrinal Dutta*[7]

*1Photovoltaic Metrology Section, Advanced Materials and Device Metrology Division, CSIR-National Physical Laboratory, New Delhi 110012, India;*

*2Academy of Scientific and Innovative Research (AcSIR), Ghaziabad, Uttar Pradesh 201002, India;*

*3 International Center for Materials Nanoarchitectonics (MANA), National Institute for Materials Science, 1-1 Namiki, Tsukuba, 305-0044, Japan;*

*4 Faculty of Petroleum and Mining Engineering, Science and Mathematics Department, Suez University, Suez 43518, Egypt;*

*5 Department of Physics, Indian Institute of Technology Roorkee, Roorkee-247667, Uttarakhand, India;*

*6 Centre of Nanotechnology, Indian Institute of Technology Roorkee, Roorkee-247667, Uttarakhand, India;*

*7 National Institute of Solar Energy, Gurgaon 122003, Haryana, India*

*Corresponding Author E-mail: duttamrinal@nise.res.in, Orcid id: 0000-0003-2747-329X*

## Abstract:

A new method of introducing nanopores with spongy morphology during fabrication of size and pitch controlled flexible silicon microwires (SiMWs) in wafer scale is presented using nanosphere lithography technique in addition to metal catalyzed electroless etching technique by varying concentration of oxidant and introducing surfactant or co-solvents to the etching solution. For achieving self-assembled monolayer closed-pack pattern of $SiO_2$ microparticles in wafer-scale simple spin coating process was used. The effect of variation of the etchant, oxidant and surfactant on the morphology and optical properties of SiMWs were studied. By simply controlling the diameter of $SiO_2$ microparticles and concentration of $H_2O_2$ the size of the MWs as well as the introduction of pores could be controlled in wafer-scale. Average reflectance suppressed to below 8% in the broad spectral range of 400-800 nm for these porous, spongy and flexible SiMW arrays in wafer scale. The mechanism behind this formation of spongy, porous and flexible nature of the SiMWs is also demonstrated for a better understanding of the etching process.

## Introduction

One dimensional microstructure coupled with porous nature have shown great promise towards the production of flexible and biodegradable electronics to be used in flexible solar cells, energy storage devices, biomedical sensors and printed circuits to be implanted in human beings.[1-3] Porous silicon (Si) has shown its potential in photovoltaics, light emitters, chemical sensors, micromachining, bone growth material etc.[4-6] Silicon in flexible and porous microwire (MW) and nanowire (NW) configurations has great potential to be used in next generation flexible Li-ion batteries, super capacitors, Li-ion capacitors, solar cells, fuel cells, photodetectors etc.[7-10] Flexible SiNW arrays in tilted or curved form have shown potential for the formation of homogeneous and continuous electrode that has remained a problem for SiNW arrays based solar cells.[11] Studies already proved that enhanced omni-directional antireflection could be achieved from size and pitch-controlled Si wire arrays over a broad range of solar light.[12-13] Flexible spongy SiMW arrays with controlled spacing could be an excellent anode material for flexible Li ion batteries and Li ion capacitors to accommodate the volume expansion of Si material during Li ion intercalation/de-intercalation to avoid pulverization of anode.[14] A vast effort has been directed in the last decade to fabricate SiMW structures in a controlled fashion using different techniques such as chemical vapor deposition (CVD), reactive ion etching (RIE) in addition to lithography, block co-polymer, interference lithography, nanoimprint lithography, nanosphere lithography (NSL) followed by metal catalyzed electroless etching (MCEE) etc.[15-21] Among the above-mentioned techniques NSL has been popularized due to its simplicity, cost effectiveness, accuracy and possibility of wafer-scale growth. Another potential fabrication technique is nanoimprint lithography. Its high accuracy and large area growth potential has made it one of the possible techniques. However, the major constraints lie in the use of time consuming and costly lithographic procedure to first fabricate master mold with a designated surface pattern in wafer-scale. Thus, NSL has got much attention worldwide compare to nanoimprint lithography technique to fabricate diameter-controlled Si wire from nanowire array to microwire array geometry just by forming a monolayer of different diameter nanosphere or microsphere on the wafer surface to act as etching mask.

In the MCEE process depending on the concentration of the oxidation agent, Si surface could become very rough, porous or textured in the form of vertical Si nanowire (SiNW) arrays.[22] Again, depending on the relative concentration of HF and $H_2O_2$ in solution, crystallographic orientation and doping type of the wafers, the MCEE process could produce rough and porous surface or tilted SiNWs.[23-25] Earlier studies show that addition of ethanol as a wetting agent and surfactant could lower the etching rate with change in the morphology from porous NW with porous surface (PS) to PS for Si in specific resistivity range and further addition of ethanol (EtOH) produces polished surface.[26] Kim et al. have shown the production of tilted SiNW arrays by using different co-solvents or wetting agents by using MCEE technique.[27] These previous investigations and results suggest that fabrication of size and space controlled flexible spongy SiMW arrays in wafer scale could provide a path to achieve flexible microelectronics to be used in fabrication of flexible biomedical sensors, flexible energy storage devices and solar cells.

Up to now, a lot of efforts have been devoted towards the growth of SiNW and SiMW arrays either by using the MCEE method or by using the NSL technique followed by MCEE. However, fabrication of porous SiMW and influence of the composition of etching solution have not been yet reported. Here, we report for the first time the formation of porous flexible SiMW arrays in wafer-scale using NSL to form an etching mask and chemical etching in a $HF-H_2O_2$ solution in presence of ethanol (EtOH). Through this work, the formation mechanism of flexible MW morphology with pores and cracks on the surface wall has been proposed and correlated to the effect of various fabrication parameters. The effects of etching in various conditions on the optical properties of these SiMW arrays have also been studied. Through this same fabrication method different size and space controlled porous Si wires could be fabricated subjected to the use of different size silica micro and nano particles and their etching by inductively-coupled plasma reactive ion etching (ICP-RIE) during NSL technique.

**Experimental**

**Materials:**

Aqueous solution of monodispersed $SiO_2$ microparticles of average size of 1.1 micron was purchased from Sigma-Aldrich. N, N-dimethylformamide (DMF), HF (48%), $H_2O_2$ (30%), $HNO_3$ (60%) and ethanol (EtOH) were purchased from Sigma Aldrich and used as received without any further purification.

**Fabrication of flexible porous SiMW arrays:**

The monodisperse $SiO_2$ microspheres aqueous solution was dried to obtain dry microparticles in powder form. These dry microparticles were dispersed in N, N-dimethylformamide (DMF) at an optimized concentration and sonicated for 2 h to produce complete dispersion of the microparticles. Si MW arrays were fabricated on 2-inch 300 μm thick p-type Si wafers of (100) orientation and resistivity of 5-10 Ω-cm. The surface of the Si wafers was made hydrophilic by following the procedure as stated below. The Si wafers were first cleaned with ultrasonication in acetone (5 min), IPA (5 min) and deionized (D.I.) water (5 min). Then immersed in boiling piranha solution (3:1 mixture of $H_2SO_4$ and $H_2O_2$) for 10 min and thoroughly rinsed with D.I. water 2-3 times. Then dipped in 2% HF solution for 5 min and rinsed with D. I. water for 2-3 times. After this the Si wafers were immersed in a 5:1:1 volume ratio mixture of D. I. water, $H_2O_2$ and $NH_4OH$ at a certain temperature for 45 min. This treatment made the cleaned wafers surface hydrophilic, and the residual water on the surface was immediately removed with blowing air (Figure 1A). Then $SiO_2$ microparticles were spin-coated in air at fixed spinning speed for 100 s after dropping 15 μL of solution on the wafer surface (Figure 1B). Spin coating formed a closed-pack monolayer of $SiO_2$ microparticles on the Si wafers surface (Figure 1B). The non-close packed structure of $SiO_2$ particles was fabricated on silicon surface following the plasma chemical etching by ICP-RIE technique for a specific time to reduce the size of the particles (Figure 1C). For this purpose, plasma was created by using $CF_4$ gas with a flow rate of 20 SCCM at RF power of 200 W and pressure of 2 Pa keeping substrate temperatures close to room temperature. The said plasma can only etch the $SiO_2$ with an etching rate of 45 nm/min and with a negligible etching of silicon underneath.

In the next step, a thin Ag coating (~ 40 nm) over the non-closed-pack monolayer of microparticles was achieved by the electron beam evaporation system (Figure 1D). During metal deposition, Ag filled through the gap among the microparticles created by the ICP-RIE process and coated on bare Si surface. To obtain an Ag mask pattern where the hole area is defined by the reduced size of microparticles after the RIE process, ultrasonication of the Si wafers was done to remove the microparticles from the Si surface. For the fabrication of Si microwires, these Si wafers were dipped in a solution with a volume ratio of HF: $H_2O_2$: EtOH = 10:1:1 (will be referred as solution A throughout the paper) for different time intervals of 15 min to 45 min so as to vary the length of SiMWs from 18 μm to 45 μm (Figure 1E) and more prolonged etching (more than 45 min) starts breaking of the SiMWs. To check the effect of etchant, oxidant and surfactant on the growth of microwires, concentration of HF, $H_2O_2$ and EtOH were varied by making double and half of the concentration that of solution A by keeping all other parameters same that fabricated SiMW arrays of average length of 34 μm for etching in solution A. After etching, all the wafers were thoroughly washed with D. I. water and dipped in concentrated (60%) and diluted $HNO_3$ for 5 min one after another to remove Ag from Si surface (Figure 1F). Finally, these wafers were rinsed with D. I. water and dried by air flow.

**Characterization:**

Microstructural characterization was done using a high-resolution field-emission scanning electron microscope (FESEM, Hitachi S-4800). Micro-Raman scattering measurements were performed with a 100× objective and a 532-nm excitation light. The excitation power was set at about 0.03 mW to avoid local heating effects caused by the excitation laser. The spectral resolution of all data was about 0.3 cm$^{-1}$. A UV–vis-NIR spectrophotometer (Perkin Elmer, Lambda 1050) was used to record UV–vis-NIR reflectance spectra in the wavelength range of 220–2000 nm employing the integrating sphere module.

**Results and discussion:**

Fabrication of uniform diameter MWs in wafer-scale with controlled pitch by nanosphere lithography technique needs the formation of a monolayer closed pack structure of the silica microparticles in wafer-scale. This monolayer closed pack structure formation of silica microparticles was obtained by simple spin coating technique. The formation of such a monolayer closed pack structure was achieved in wafer-scale. Figure 2(A) shows the image of a 2-inch Si wafer coated with silica microparticles. Figures 2(B) and 3(A) show the corresponding

FESEM images of the closed pack monolayer structure of silica microparticles with an average diameter of 1.1 µm that support formation of closed-pack structure in long-range order. Uniform size reduction with a round shape and average diameter of 1.01 µm of the silica microparticles, that determine the diameter of the MWs and also the pitch, was obtained by RIE as shown in Figures 3 (B) and (C). Prior to fabrication of Si MW array by MCEE technique, formation of uniform metal (e.g. Ag) mask by removing non-closed-pack monolayer of silica microparticles is shown in Figure 3 (D).

Earlier studies show that morphology of the textured porous surface of Si obtained by electrochemical etching, consists of not only pores but also cracks affected by fabrication parameters.[28-29] Similar observations were made during the fabrication of flexible and porous SiMWs in our present study. Figure 4 shows the FESEM images of the flexible spongy tipped SiMW arrays obtained by varying etching time from 15 min to 30 min. The average diameter determined by the reduced size of silica microparticles after the RIE process is estimated as ~ 1.01 µm and depending on the chemical etching time in solution A, the average height of the MWs are varied from 18 µm to 34 µm for varying the etching time from 15 min to 30 min (Figure 5 A and B). Figure 4 (C) and (D) shows the spongy morphology of the tips of the MWs. From Figure 4 (C) and (D) it is clear that the tips and upper end of these MWs are spongy nature. However, this type of morphology could not be observed on the lower sidewalls of the MWs. Figure 6 shows the effect of longer etching in solution A on the surface walls of MWs. With the increase of etching time macropores are introduced on the sidewall of the MWs and sometimes fractures are also created. In a long run these fractures became the cause of breaking of MWs. However, relatively smooth sidewalls were observed for shorter etching time up to 30 min (Figure 6 A). With increase of etching time up to 45 min cracks and fractures are introduced on the sidewalls of the MWs and average height increased up to 44 µm as shown in Figure 6 (B) and 5 (C). For more prolonged time of etching up to 60 min, these fractures cause breaking of the MWs that could be evident from Figure 6 (C) and (D). Figure 7 (A)-(C) show the effect of variation of $H_2O_2$ concentration on the morphology of SiMWs. When the concentration of $H_2O_2$ became nearly half that of solution A (in a ratio of HF: $H_2O_2$: EtOH = 10:0.5:1) by keeping other parameters unchanged, etching for 30 min produced flexible and spongy tipped microwires. However, with increase of $H_2O_2$ concentration up to nearly double by keeping all the other parameters same (HF: $H_2O_2$: EtOH = 10:2:1), formation of nanopores with diameter in the range 35-66 nm could be observed on the sidewall of the MWs with an etching rate of 12.77 nm/sec as shown in the FESEM image (Figure 7 C). Whereas low concentration of $H_2O_2$ could not able to create any visible pore on the sidewall of the MWs though etching rate reduced to 10.82 nm/sec. Increasing the concentration of ethanol nearly double that of the solution A while keeping the other etching parameters same (HF: $H_2O_2$: EtOH = 10:1:2) shows formation of micro cracks on the sidewall of the MWs with an etching rate of 16 nm/sec. However, etching with a low concentration of ethanol (HF: $H_2O_2$: EtOH = 10:1:0.5) increases etching rate from 16 nm/sec to 20 nm/sec and no formation of such cracks were noticed. On the other hand, increasing the concentration of HF by 1.5 times that of the solution A creates MWs with rough tips instead of spongy tips and relatively smoother sidewalls as compare to previous observations (Figure 8 A and B).

Since this is an electroless chemical etching process, the effect of local electrochemical conditions on the morphology of etched Si could be understood by taking into consideration of molar ratio of $\rho = [HF]/\{[HF]+[H_2O_2]+[EtOH]\}$ as the concentration of $[H_2O_2]$ is equivalent to current density of electrochemical processes and concentration of [HF] rules over the surface chemistry.[23] Ethanol in this process acts as a surfactant and reduces surface energy. Throughout the work this molar ration has been kept in the range of 70%< $\rho$ < 100% as previous works on electrochemical etching of Si surface in the presence of Ag shows that only within this region Ag nanoparticles produces straight cylindrical pore.[23]

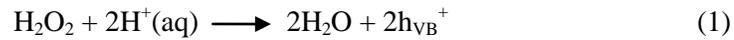

$$H_2O_2 + 2H^+(aq) \longrightarrow 2H_2O + 2h_{VB}^+ \qquad (1)$$

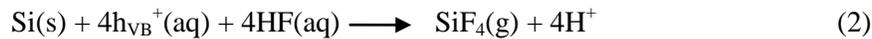

$$Si(s) + 4h_{VB}^+(aq) + 4HF(aq) \longrightarrow SiF_4(g) + 4H^+ \qquad (2)$$

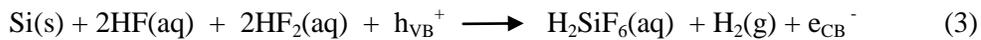

$$Si(s) + 2HF(aq) + 2HF_2(aq) + h_{VB}^+ \longrightarrow H_2SiF_6(aq) + H_2(g) + e_{CB}^- \qquad (3)$$

In the process of etching $H_2O_2$ decomposes to inject holes ($h_{VB}^+$) (equ 1).[30] After formation these holes energize HF molecules to react with Si atoms at the surface and remove as $SiF_4$ (equ 2). $SiF_4$ dissolved in HF to form $H_2SiF_6$ (equ 3).[31] During the initial period of etching in solution A the reaction is HF dominated and all holes generated at Ag/Si interfaces are not consumed. These unconsumed holes diffuse to the parts of Si surface where Ag was not deposited. Hence spongy porous Si formed at the tips of the SiMWs. With the passage of time concentration of $H_2O_2$ reduces (equ 2) and the etching process starts to dominate by the concentration of $H_2O_2$ as there was a sufficient amount of HF. In this stage most of the holes produced at the Ag/Si surface are consumed. This phenomenon is well reflected in etching of patterned Si wafer in solution A as shown in Figure 4 except the occurrence of some fractures for a long etching time (Figure 6). In presence of excess HF for etching solution containing 1.5 times of HF concentration than that of solution A spongy porous tips were not formed, rather the top surface of MWs became rough (Figure 8). For etching in a solution with a double concentration of $H_2O_2$, the reaction was dominated by the concentration of HF in the solution. Etching in this solution for the same time period resulted in the formation of pores on the side walls of the MWs in addition to the spongy tips that could be attributed to the generation of excess holes due to higher concentration of $H_2O_2$ in the solution and diffusion of these excess holes towards the wall of the SiMWs.

In the course of the reaction process ethanol evaporates and the equilibrium between surface tension and pressure disrupts. The surface pressure becomes greater, which creates structural instability. The relation between this increased surface pressure with surface tension relating to pore size could be realized using the Young-Laplace equation,[32]

$$\Delta p = - 2\gamma/r \qquad (1)$$

Where $\Delta p$ is the differential pressure, $\gamma$ is the surface tension of the solution, and r is the pore radius.

This equation suggests that through the increase in pore size, an increase in surface pressure could be compensated for, leading to formation of cracks for the long time reaction process. The cause of bending of the MWs could be realized using the relations by analyzing the model involved in deflection engineering of a uniformly loaded cantilever (Figure 8C).[33]

$$\delta_B = - q\, l^4/8EI \qquad (2)$$

$$\Phi_B = - q\, l^3/6EI \qquad (3)$$

Where $\delta_B$ is the bending deflection of SiMW, $\Phi_B$ is the angle of the bending, $q$ is the transverse pressure received, $l$ is the height of the SiMW, and $E$ is modulus of elasticity and I is the area moment of inertia of SiMW cross section. Here, EI is kept constant. From equations (2) and (3) it is clear that with increase of SiMWs length ($l$), $\delta_B$ and $\Phi_B$ increases for the same transverse pressure (P). Under transverse pressure, bending of the SiMWs leads to structure deformation and forms larger cracks for longer etching time. In this situation the extrusion force in pores and cracks leads to breakage of SiMWs for prolonged etching time.

Raman scattering measurements were further performed on these SiMWs at room temperature to investigate the effect of tensile stress in addition to the phonon confinement effect in porous structure depending upon the growth conditions as shown in Figure 9. The peak centered at 520.1 cm$^{-1}$ corresponds to the optical phonon peak of bulk-Si (Figure 9A). Depending on the growth conditions Raman peak shift compared to the bulk-Si optical phonon peak position could be observed in all samples. With increase of etching time low frequency shift occurred that increases with etching time up to 45 min. However, for etching time 60 min relaxation in the red shift of Raman peak could be noted from Figure 9 (A). In a similar fashion Raman peak shows slight asymmetric broadening towards the lower wavenumber side. Increasing $H_2O_2$ concentration by double introduced pores on the side wall of the MWs that reflected in higher low frequency shift of Raman peak upto 518.75 cm$^{-1}$ compared to the MWs etched in a solution with half of $H_2O_2$ concentration (Figure 9 (B)). Larger red shift observed for the MWs etched in solution with double concentration of ethanol compared to that of solution A results from the introduction of pores and fractures in MWs. With change of HF concentration than that of solution A also introduces a red-shift of Raman peak as shown in Figure 9 (D). However, for the MWs with double concentration of HF than that of solution A shows some relaxation in stress level that could be attributed to more aligned MW arrays and non-porous structure of these MWs as observed in FESEM images. All these red-shifts observed for MWs are within 2 cm$^{-1}$ with respect to optical phonon peak of bulk-Si and could be attributed to the increase of tensile stress with increase of etching time and formation of pores on the sidewalls of the MWs.[34-37] The behavior of such Raman peak shift and broadening of the sharp optical phonon peak could be explained by using the theory based on phonon confinement effect developed by Richter et al. and later improved by Campbell and Fauchet for microcrystalline Si.[38-39] One of our authors has observed the phonon confinement effect for Si nanostructures.[40-41] However, in present study there are some contradictions in comparison with the assumptions made in formulation of this theory. For Si MWs studied at present it is unknown about the chemistry of surface states and also stress inhomogeneity may prevail along the axis of the MWs.[42] However, a detailed theoretical and experimental investigation on this effect of strain and pores by studying the Raman scattering is beyond the scope of the present work and hence will be presented elsewhere.

Again, for instance, there is growing interest in the use of Si nanostructure with an optimized period and size for solar cells. Spinelli et al. has shown that Si nanopillars with certain size and pitch can enhance trapping of incident light over a broad wavelength range that improves omnidirectional antireflection.[12] Furthermore, the study by Sang Eon et al. demonstrated the significance of Si nanostructure size (or Si filling fraction) in producing diffraction of transmitted light for light absorption enhancement.[13] Figure 10 shows the hemispherical reflectance measurements performed over the porous SiMW arrays with different lengths obtained by varying etching time and concentration of etchant, oxidant and surfactant in comparison to that of a polished Si wafer over a broad spectral range of 200-1400 nm. The sharp transition in the range 1000-1200 nm corresponds to the band edge of Si.[43] The reflectance of the polished planar Si wafer varies from 66% - 42% in the range of 300–1000 nm, whereas the reflectance of the SiMW arrays noticeably suppressed over a broad spectral wavelength range from 200–1000 nm. Due to the light scattering and sub-wavelength light trapping among the SiMWs, the average reflectance of SiMW arrays could be suppressed to less than 8% in the 400-800 nm range compared to an average reflection of 45% from polished surface of Si. Figure 10 (A) shows the reflectance spectra of the SiMWs of different lengths. With increase in average length of the MWs from 18 μm to 45 μm average reflectance decreases from 23.5% to 16.4% in the 400-800 nm range. With the breaking of the MWs reflectance again starts to increase. Figure 10(B) shows the effect of variation of solution composition on the reflectance of the SiMW arrays. Increase in ethanol concentration increases the reflection which could be ascribed to the reduction in etching rate with increase of ethanol concentration.[26] Increase in $H_2O_2$ concentration decreases the reflection with an average reflection of 7% in the range of 400-800 nm. This reduced reflection could be ascribed to the absorption of light in the nanopores formed on the SiMWs walls during the etching process as observed in Figure 7 (C) in addition to the sub-wavelength light trapping and light scattering in SiMW arrays. However, increase in HF concentration increases the reflection due to reduced etching rate and relatively smooth tips and walls as evident during microscopic observations (Figure 8).

In conclusion, size and pitch controlled flexible and spongy SiMW arrays are fabricated in wafer scale with different length by varying the etching time. NSL technique has been used to fabricate self-assembled monolayer non-closed pack structure of $SiO_2$ microspheres mask pattern in wafer scale by using simple spin coating in addition to ICP-RIE for deposition of Ag metal mask pattern. Nanopores were introduced by controlling the oxidant concentration. Spongy porous morphology could be tuned by simply changing the concentration of etchant. Based on the morphological observation a growth mechanism is proposed to understand the formation of flexible, spongy and porous structure SiMW arrays. Effect of different growth conditions such as etching time, concentration of etchant, oxidant and surfactant on the UV-Vis-NIR reflection property of the SiMW arrays was studied. Reflectance suppressed to as low as 7% for the porous spongy flexible SiMW arrays in a broad spectral range of 400-800 nm.

## Authors Contributions

M. D. conceived the project. Anjali and M. D. conceptualized the work, synthesized and characterized the samples, D. R. and A. S. performed the ICP-RIE process and done FESEM characterization, N. F., M. A. and W. J. performed FESEM and Raman Characterizations. The manuscript was written through contributions of all authors. All authors have given approval to the final version of the manuscript.

## Notes

The authors declare no competing financial interest.

## Acknowledgments

The first author wishes to thank Council of Scientific and Industrial Research (CSIR) for providing the doctoral research fellowship. This work was supported in part by DST-SERB Ramanujan Fellowship (File no. SB/S2/RJN-077/2017) and DST-SERB (Project no: ECR/2017/001050).

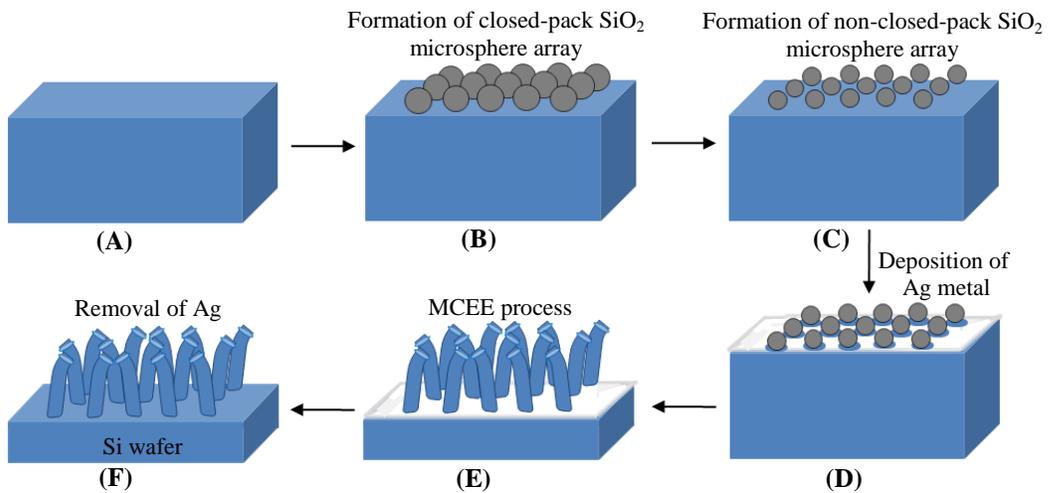

**Figure 1.** Schematic of the SiMW array fabrication process using nanosphere lithography followed by to MCEE technique.

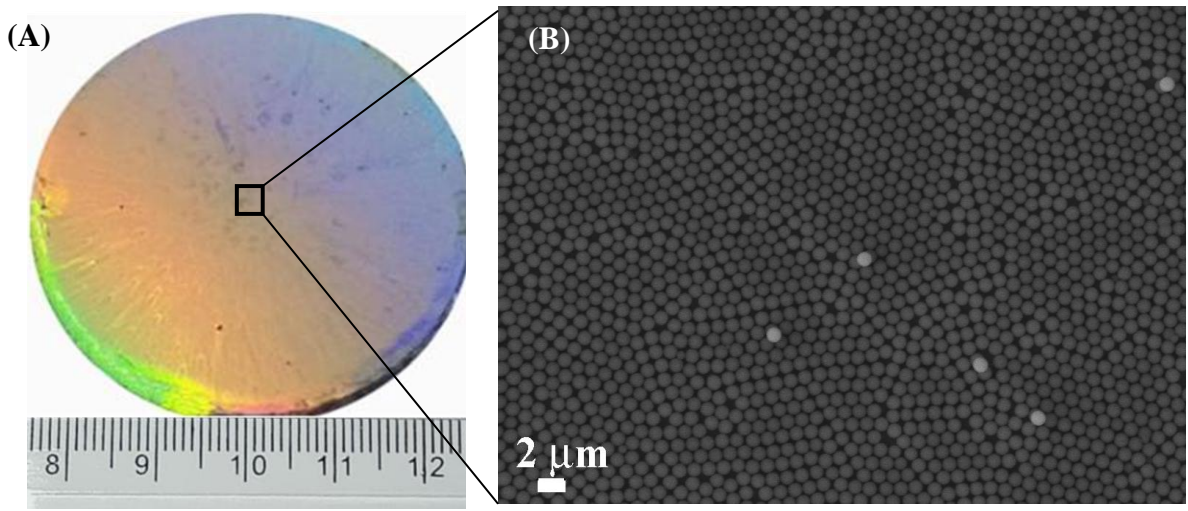

**Figure 2.** (A) closed pack monolayer pattern of $SiO_2$ microparticles on 2-inch Si wafer under white light illumination, (B) SEM image of the closed pack monolayer of $SiO_2$ microparticles as shown in (A).

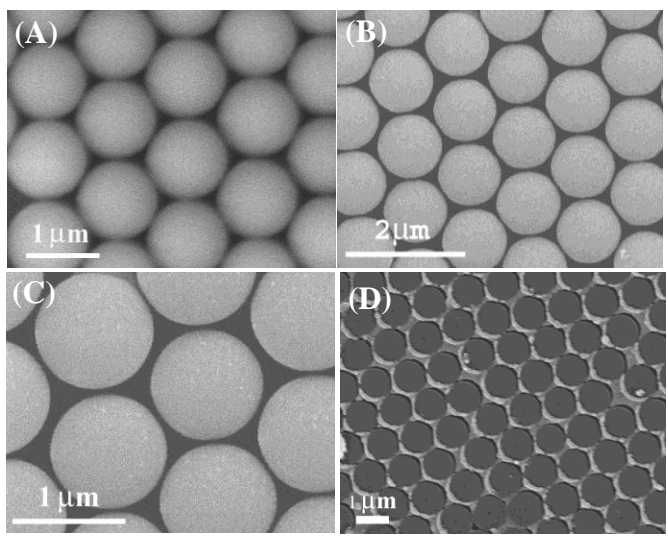

**Figure 3.** (A) High magnification SEM image of closed pack monolayer of silica microparticles, (B) and (C) SEM images of the non closed pack monolayer of silica particles obtained after RIE process. (D) Periodic patterned Ag mask formed by removing non closed pack silica microparticles.

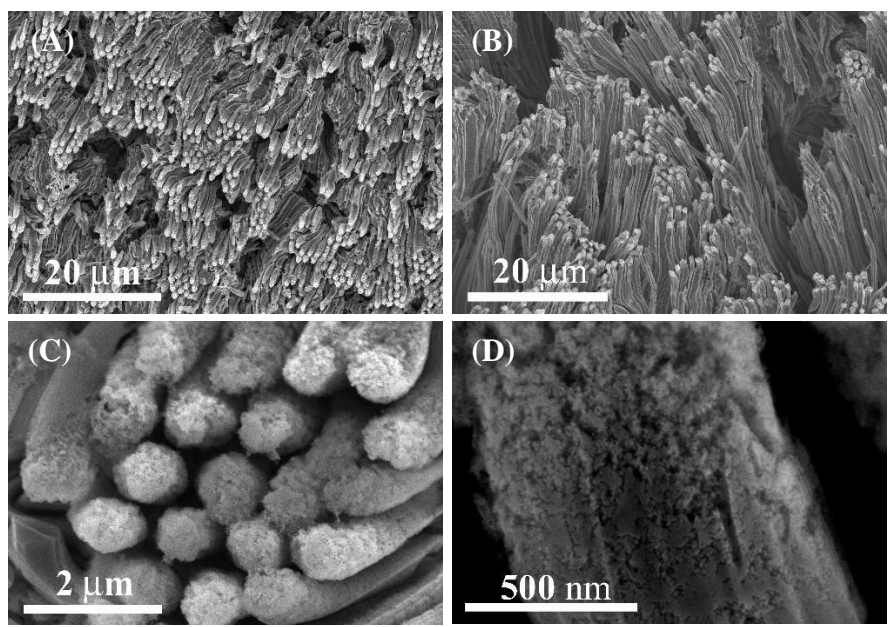

**Figure 4.** FESEM images (top view) of the SiMWs obtained by etching in solution A for (A) 15 min, (B) 30 min. (C) High magnification top view FESEM image of the spongy tips of the MWs. (D) Upper side view of the spongy tip of a typical MW.

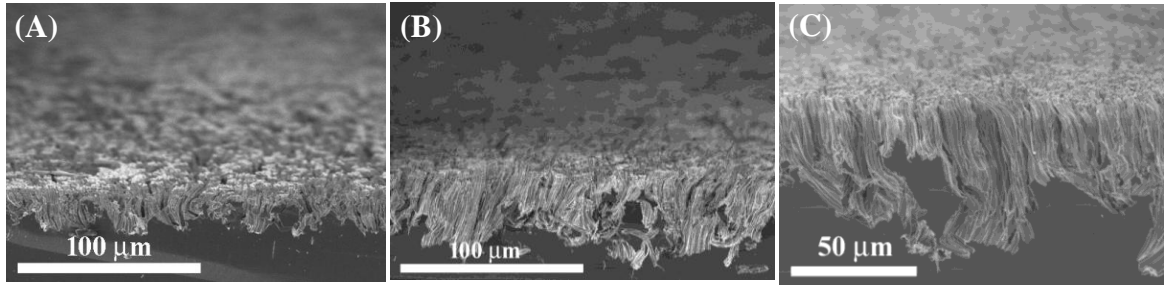

**Figure 5.** $10^0$ tilted cross FESEM images of the SiMWs obtained by etching in solution A for (A) 15 min, (B) 30 min and (C) 45 min.

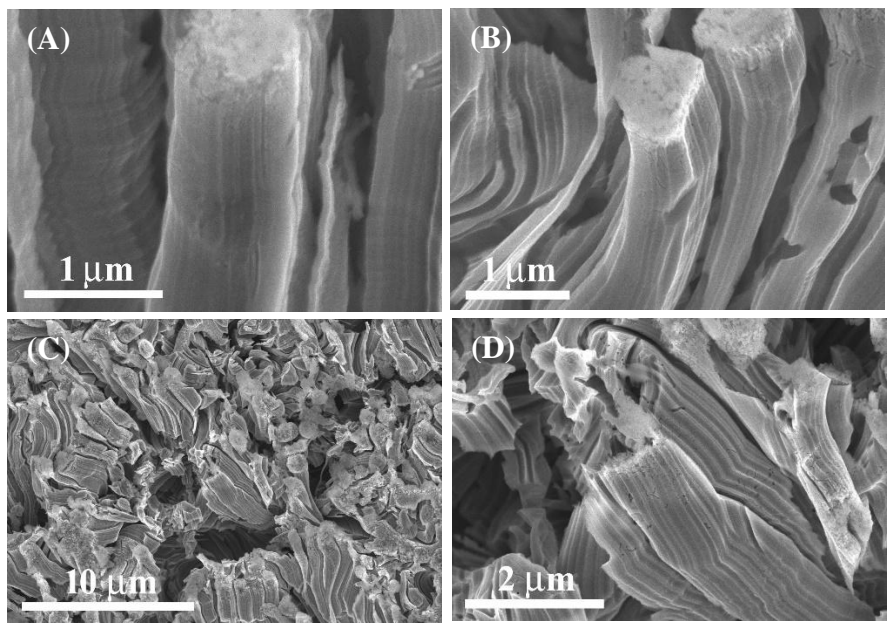

**Figure 6.** (A) FESEM image of a typical MW etched for 30 min in solution A, (B) FESEM image of MWs showing formation of cracks on the surface of MWs for etching time of 45 min. (C) and (D) FESEM image of the broken MWs formed by etching for 60 min in solution A.

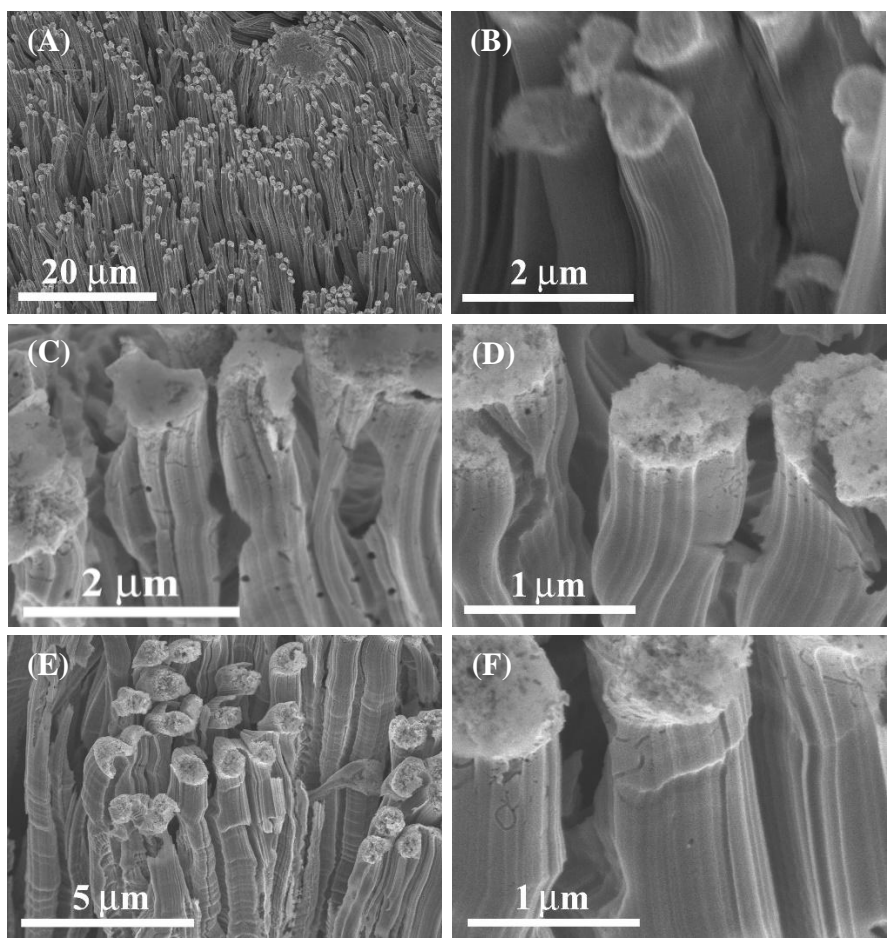

**Figure 7.** (A) and (B) FESEM images of SiMW arrays obtained by etching for 30 min with half concentration of $H_2O_2$ in solution A keeping all other parameters constant, (C) FESEM image of the SiMWs obtained by etching with double concentration of $H_2O_2$ in solution A keeping all other parameters constant, (D) FEEM image of SiMWs obtained by etching with double concentration of ethanol in solution A keeping all other parameters constant. (E) and (F) FESEM images of the SiMWs arrays obtained by etching in solution A with half the concentration of ethanol keeping all other parameters constant.

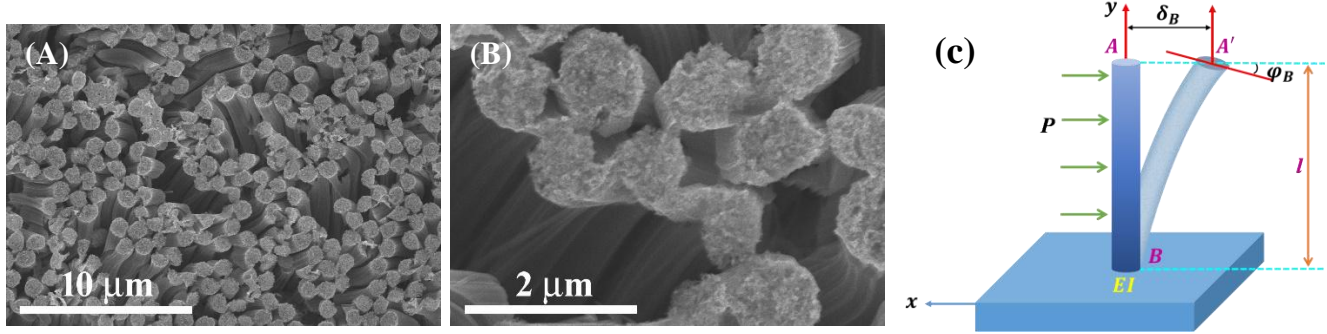

**Figure 8.** (A) and (B) FESEM images of SiMW arrays obtained by etching in a solution with 1.5 times concentration of HF of solution A keeping all other etching parameters unchanged. (C) Model to understand the mechanism behind SiMWs bending, pore and fracture formation.

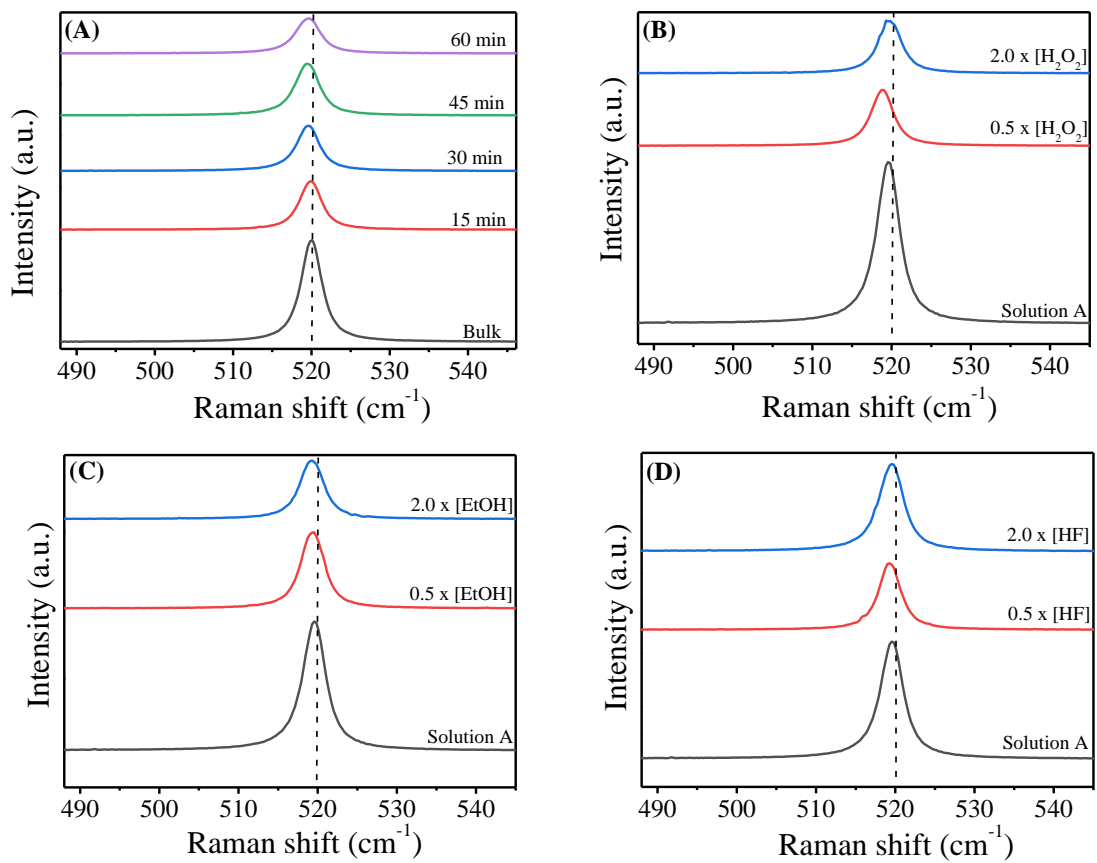

**Figure 9.** Raman spectra of (A) SiMWs fabricated by etching in solution A for different time. (B), (C) and (D) Raman spectra of the SiMWs obtained by etching in solution A for 30 min with the variation in concentration of oxidant, surfactant and etchant.

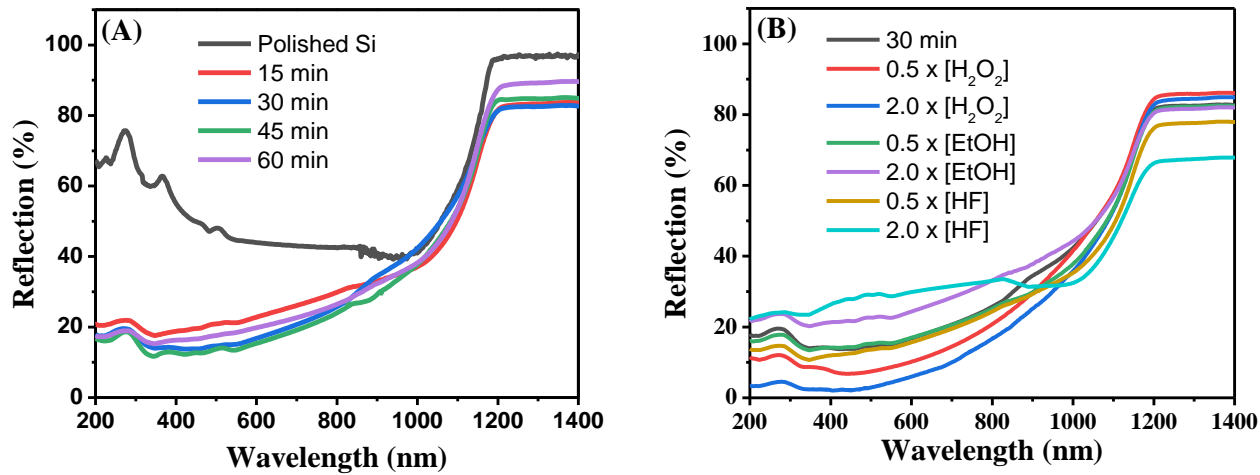

**Figure 10.** (A) UV-Vis-NIR reflection spectra of planar polished Si wafer and SiMW arrays of different length obtained by etching in solution A for different time and (B) UV-Vis-NIR reflection spectra of the SiMWs obtained by etching in solution A for 30 min with the variation in concentration of oxidant, surfactant and etchant.